\documentstyle[preprint,tighten,aps]{revtex}
\draft
\begin{document}
\title{Effects of in-chain and off-chain substitutions \\
on spin fluctuations in the spin-Peierls compound CuGeO$_3$}
\author{\bf P. Lemmens$^a$, M. Fischer$^a$,  G. G\"untherodt$^a$, \\
C. Gros$^b$,\\ 
P.G.J. van Dongen$^c$, \\
M. Weiden$^d$, W. Richter$^d$, C. Geibel$^d$, F. Steglich$^d$} 
\address{
$^a$ 2. Physikalisches Institut, RWTH-Aachen, 52056 Aachen, 
Germany\\
$^b$ Institut f\"ur Physik, Universit\"at Dortmund, 44221 Dortmund,
Germany\\  
$^c$ Theoretische Physik III, Universit\"at Augsburg, 86135 
Augsburg,
Germany\\
$^d$ FB Technische Physik, TH-Darmstadt, 64289 Darmstadt, 
Germany\\}
\date{Febr., 20th. 1997}
\maketitle
\begin{abstract}
The effect of in-chain and off-chain 
substitutions on 1D spin fluctuations in the spin-Peierls compound 
CuGeO$_3$ has been studied using Raman scattering in order to
understand the interplay between defect induced states,
enhanced spin-spin correlations
and the ground state of low dimensional systems. 
In-chain and off-chain substitutions quench the spin-Peierls 
state and induce 3D antiferromagnetic order at T$\leq$5~K. 
Consequently a suppression 
of a 1D gap-induced mode as well as a constant intensity 
of a spinon continuum are observed at low temperatures. 
A 3D two-magnon density 
of states now gradually extends to higher temperatures T$\leq$60~K
compared with pure CuGeO$_3$. 
This effect is more pronounced in the 
case of off-chain substitutions (Si) for which a N\'eel state 
occurs over a larger substitution range, starting 
at very low concentrations. Besides, additional 
low energy excitations are induced. 
These effects, i.e. the shift of a dimensional crossover to 
higher temperatures are due to an enhancement of 
the spin-spin correlations induced by 
a small amount of substitutions. The results are compared
with recent Monte Carlo studies on substituted spin ladders, 
pointing to a similar
instability of coupled, dimerized spin chains and spin ladders 
upon substitution.

\end{abstract}

\section{INTRODUCTION}
Quantum effects in low-dimensional antiferromagnetic spin systems 
have attracted an intensive theoretical and experimental 
interest in recent years. 
One of these effects is the formation of a continuum
of spin fluctuations in one-dimensional (1D) spin-1/2 chains 
\cite{Mueller}.  
The magnetic excitations of the uniform s=1/2 Heisenberg 
antiferromagnetic 
chain, can be derived using the Bethe Ansatz. They are
dominated by strong fluctuations and
consist of delocalized or unbound soliton-like spinon excitations 
prohibiting a magnetically 
ordered state. The spinon continuum is enclosed by
an upper and lower dispersion relation 
of the form E$^2_q$=$\pi$J$\vert$sin(q$_c$/2)$\vert$ and 
E$^1_q$=($\pi$/2)J$\vert$sin(q)$\vert$, respectively
\cite{Mueller,Cloizeau}. The existence of this continuum was
verified in neutron experiments on the compound KCuF$_3$ 
\cite{KCuF3}
and later in CuGeO$_3$ \cite{Arai}. 

Of particular importance in this context is the spin-Peierls (SP)
transition 
leading to a singlet ground state of the formed dimers and the opening 
of a spin gap near the lower boundary of the continuum.
The modification of the excitation scheme in the 
dimerized state is theoretically not completely settled. The change 
under dimerization from spinons with s=1/2 to
excitations with s=1 should lead to a redistribution 
of density of states from the spinon continuum into its lower
dispersing limit. 
This may be understood as a localizing, pair-binding effect of the
excitation gap \cite{Uhrig} 
due to the enhanced spinon-spinon interaction in the gapped state.
The spin-spin correlation length changes from an algebraic to
an exponential decay.

The compound CuGeO$_3$ initially appeared as an ideal spin-Peierls
compound with a high T$_{SP}$=14~K and a stable lattice
which allows for several types of substitutions 
\cite{Hase,Nishi,Raman,Boucher}. 
Later, however, evidence
appeared that its behavior is complicated due to competing 
exchange interactions \cite{Castilla,Buech}. 
The strong frustration of the spin system may drive or at least stabilize
the dimerization transition in this compound. 
CuGeO$_3$ consists of chains of spin-1/2 Cu$^{2+}$ ions 
along the c-axis coupled by antiferromagnetic 98$^\circ$-
superexchange 
through oxygen orbitals \cite{Hase,Nishi,Braden}. 
This special superexchange geometry   
causes a strong sensitivity of the exchange on changes of the 
Cu-O-Cu bond angle.
A bond angle close to the critical 90$^\circ$ leads to a 
comparatively small antiferromagnetic 
nearest neighbor (nn) exchange, J$_{c1}$, and a frustrated 
next-nearest neighbor (nnn) 
exchange, J$_{c2}$, that strongly depends on the local 
surrounding of the oxygen \cite{Khomskii}.

The exchange along the Cu-O-Cu chains can be modeled 
by a 1D-Hamiltonian including the competing nn and 
nnn interactions ($\sim\alpha$) and the dimerization $\delta$
\cite{Castilla}: 
\begin{center}
\begin{equation}
H = J_{c1}\sum_i [ (1+\delta(-1)^i){\bf S}_i\cdot{\bf S}_{i+1}
            +    \alpha       {\bf S}_i\cdot{\bf S}_{i+2}
            ]~.
\label{H}
\end{equation}
\end{center}
Using neutron scattering results \cite{Nishi} the following exchange 
constants were derived:
J$_{c1}$=150~K, J$_{c2}$$\approx$36~K, 
with the frustration parameter $ \alpha$=J$_{c2}$/J$_{c1}$=0.24-0.36 
close to or larger than a critical $\alpha_{cr}$=0.2411 \cite{Castilla}. 
For increasing $\alpha$$\geq$$\alpha_{cr}$ an exponentially small 
excitation gap exists which evolves to $\Delta$=J$_{c1}$/4 in a
valence bond solid with $\alpha$=0.5 \cite{Chitra}. 
For CuGeO$_3$ a singlet-triplet gap $\Delta_{SP}$=24-30~K
\cite{Nishi,Martin} 
and a successive gap of similar size, separating the 
singlet-triplet excitation from a continuum of unbound spinons, 
were found in neutron scattering below T$_{SP}$ \cite{Ain}.  
Additionally, well behaved quasi-magnon branches were observed 
below 
T$_{SP}$ \cite{Nishi} that clearly mark the change of the magnetic 
response from the uniform to the dimerized state. 

Due to the nonnegligible inter-chain exchange with 
J$_{b}$=0.1J$_{c1}$ 
\cite{Nishi} comparable to and competing with the frustrated 
nnn-exchange the compound develops 3D magnetic 
correlations below 15~K. 
This may also be the origin of the N\'eel ordering observed for
T$\leq$5~K 
in substituted samples if the coherent singlet ground state is locally 
destroyed. 
Independent of the type of substitution, T$_N$ 
is always of the order of 5~K and well separated from the 
spin-Peierls temperature of weakly substituted samples. 
In an intermediate concentration range a coexistence of  
T$_{SP}$ and T$_N$ is observed \cite{Regn2,Hase2,Fukuyama}.

Raman experiments on CuGeO$_3$ proved to be a versatile 
tool to investigate magnetic excitations due to their high resolution 
and large scattering cross section.  
Spin-Peierls active phonon 
modes together with a gap-induced mode at 30~cm$^{-1}$ 
($\approx$ 2$\Delta$$_{SP}$) and 
a two-magnon density of states with a cutoff at 226~cm$^{-1}$ were 
detected for T$<$T$_{SP}$ \cite{Raman}. 
More interesting is the broad continuum 
from 100-500~cm$^{-1}$ observed for T$>$T$_{SP}$ and the strong
redistribution of its intensity into the two-magnon signal 
observed in the dimerized state \cite{Loosdrecht,Thomsen,Lemmens}.

As shown recently, the continuum and the gap-induced mode are
described by frustration-induced Heisenberg
exchange scattering \cite{Muthu}. In these calculations a dimerization 
of the exchange constants of the Heisenberg and the Raman operator 
is taken into account as described by Eqs. 1 and 2.
The Raman operator in A$_{1g}$ symmetry with the 
dimerization $\gamma$ and the frustration $\beta$ 
is proportional to \cite{Fleury}:
\begin{center}
\begin{equation}
H_R \propto \sum_i\,[(1+\gamma(-1)^i)\,{\bf S}_i\cdot{\bf S}_{i+1}\,
+ \beta\,{\bf S}_i\cdot{\bf S}_{i+2}]~.
\label{H_R}
\end{equation}
\end{center}
A very important point to note is that in the uniform state 
($\delta=\gamma=0$), Raman intensity in the temperature 
range for negligible inter-chain coupling 
(J$_b\sim$15~K$\ll$T$<$J$\sim$150~K), 
is induced by the n.n.n. term in Eq. 1 and 2. 
When $\beta-\alpha=0$, the Raman  
Hamiltonian commutes with the Heisenberg Hamiltonian and there is 
no Raman scattering \cite{Muthu}. For lower temperatures these 
constraints are relaxed due to the inter-chain coupling. 
In this way the observed scattering continuum 
(above and below T$_{SP}$) and the gap-induced mode 
at 30~cm$^{-1}$ (below T$_{SP}$) can be modeled
by 1D-spinon excitations. The parameters used are the exchange 
coupling constant J$_{c1}$=150~K and 
the frustration $\vert$ $\beta$-$\alpha$$\vert$=0.24 
derived from neutron scattering 
and susceptibility measurements \cite{Castilla}. 
All other scattering contributions are  
negligible within a 1D-model \cite{Muthu}. 

However, the evolution to the quasi-3D two-magnon density of 
states for T$<$T$_{SP}$ 
and the related quasi-magnon dispersions 
emerging out of the spinon continuum as observed in neutron
scattering 
remain to be explained. 
This dimensional crossover from 1D to 3D upon cooling and the 
observation of a 3D magnetically ordered state in substituted samples 
at lower
temperatures (T$\leq$5~K) are pointing to an interplay of the 
spin-Peierls order parameter with 3D magnetic interactions that 
should be further clarified. 
This motivated our Raman study on in- and off-chain substituted 
single crystals of CuGeO$_3$. 

The paper is organized as follows. In section II we describe 
the experimental setup and the samples investigated. 
In section III. we present results of magnetic Raman scattering 
with its discussion in section IV. Section V. contains a summary 
of our results.

\section{EXPERIMENTAL}
We have performed Raman scattering experiments on well
characterized 
single crystals in quasi-back\-scat\-tering geometry 
with the polarization of incident and scattered light parallel to 
Cu-O chains (c-axis). In other scattering geometries no 
magnetic scattering contribution was observed. 
The experiments were performed 
with the $\lambda$=514.5-nm excitation line of an Ar-laser and a laser
power below 0.2~W/cm$^2$. The incident radiation does not
increase
the temperature of the sample by more than 1~K. The samples 
themselves are 
glued onto a copper sample holder and immersed in flowing He-
exchange gas.

Here we present results on single crystals of
(Cu$_{1-x}$,Zn$_x$)GeO$_3$ 
and Cu(Ge$_{1-y}$,Si$_y$)O$_3$ with 0$\leq$x$\leq$0.06 and
0$\leq$y$\leq$0.1 (see Table I). 
In previous studies  
the solubility range and phase homogeneity were checked for
0$\leq$x$\leq$0.1 
and 0$\leq$y$\leq$0.5 and confirmed by using SQUID-magnetometry,
X-ray 
scattering and thermal expansion measurements on both poly- and
single 
crystals \cite{Weiden1,Weiden2}. In the case of Si- substitution a  
disagreement is found between poly- and single crystals 
concerning the phase boundaries of T$_{SP}$ vs. y \cite{Weiden3}. 
This is due to a growth of single crystals close to the decomposition
temperature of the compound which shifts to lower temperature 
in Si-substituted samples.
As a result oxygen nonstoichiometry and disorder 
are expected \cite{Weiden3,Renard}. 
This is confirmed by phonon Raman scattering.
In Si-substituted single crystals several symmetry forbidden modes 
of pure CuGeO$_3$ are observed 
due to the relaxed momentum conservation. 
In addition static and temperature dependent defect modes are
observed. 
For a complete description we refer 
to a forthcoming publication \cite{Fischer}.

\section{RESULTS}
In the case of magnetic Raman scattering experiments in CuGeO$_3$ 
the
results may be divided into three temperature regimes, as first
described by 
van Loosdrecht et al. \cite{Loosdrecht}. 
At high temperatures a broad, quasi-elastic tail is visible. 
This tail is due to diffusive processes and follows roughly a 
1/$\omega$-dependence, with $\omega$ the Raman-shift 
of the experiment. At lower temperature, 15~K$\leq$T$\leq$60~K a
broad 
continuum for 100~cm$^{-1}\leq\omega\leq$500~cm$^{-1}$ is
observed 
which is assigned to the spinon continuum \cite{Loosdrecht,Muthu}.
Below 
T$_{SP}$=14K several new modes appear: three phonons at 
105~cm$^{-1}$, 370~cm$^{-1}$ and 820~cm$^{-1}$ 
(with lower intensity), a two-magnon scattering
cutoff at 226~cm$^{-1}$ and a gap-induced mode at 30~cm$^{-1}$.
We will concentrate here on the temperature range T$\leq$60~K
and the effect of substitutions on the spinon continuum and the
Peierls-transition-induced modes at 30, 226 and 370~cm$^{-1}$. 
As an example we show Raman spectra of pure CuGeO$_3$ in Fig. 1a
for temperatures between 5 and 40K.  

While the occurence of the spin-Peierls induced Raman modes 
is well established,
their intensity was not studied in detail \cite{Raman,Loosdrecht}.
However,
the intensity as function of temperature and substitution gives 
important
information about the evolution of the singlet-triplet excitations as
shown
in Fig.1b for pure CuGeO$_3$. 
While the intensity of the spinon continuum 
was determined by integrating the intensity between 
100-500~cm$^{-1}$, subtracting a linear background and the
phonons, 
the intensity of the two-magnon density of states was 
determined at the frequency of 226~cm$^{-1}$. 
At T$_{SP}$ the scattering intensity of the spinon continuum 
clearly shows a maximum as function of temperature. The following 
sharp decrease toward low temperatures is counteracted  
by the increase of the intensities at 226 and 
370~cm$^{-1}$. The gap-induced mode at 30~cm$^{-1}$ emerges 
at lower temperatures, i.e. below 11~K 
due to the evolution of a long range
coherent, dimerized state below T$_{SP}$. The competition between 
the 
1D spinon continuum, the quasi-3D two-magnon density of states 
and the Peierls-active phonon marks the proposed dimensional
crossover 
of the compound at T$_{cross}$=11~K 
induced by the spin-Peierls transition \cite{Muthu}.

T$_{cross}$ is defined here as the temperature of half maximum
intensity
of the phonon.
It should be noted that the behavior of the
two-magnon signal at 226~cm$^{-1}$ is completely different from
conventional
magnetic light scattering in 3D antiferromagnets (AF) close to T$_N$ 
or paramagnon scattering above T$_N$ \cite{Cottam}. 
In 3D AF the two-magnon scattering shows a strong broadening of
the line
shape
and shifts to lower frequencies with increasing temperature.
Additionally,
the integrated intensity shows a large increase at T$_N$ and is
therefore 
observable up to high temperatures. In CuGeO$_3$ we see 
close to T$_{SP}$ neither a shift of the two-magnon signal to lower
frequency
nor a broadening. 
The intensity of both the two-magnon signal and the phonon at 
370~cm$^{-1}$ emerge at T$\leq$T$_{SP}$ without a 
preceding energy renormalization.  
This aspect is confirmed by neutron 
scattering in the sense that magnons with a well defined dispersion
relation 
exist \cite{Nishi,Ain} only for T$\leq$T$_{SP}$. 
Questions arise concerning the character and dimensionality
of the underlying excitations. This problem will be addressed  
after discussing the results on substituted samples.

The onset of the 1D continuum in Raman scattering
at 100~cm$^{-1}$ corresponds roughly to two times the 
energy at which the continuum is observed in neutron scattering
\cite{Ain}.

However, the value of the gap-induced mode 
in Raman scattering at 30~cm$^{-1}$$\equiv$43~K is markedly 
smaller than
two 
times the singlet-triplet transition, 
$\Delta_{SP}$=24-30~K  derived from 
neutron scattering \cite{Martin,Ain}.
This points to strong attractive magnon-magnon interaction 
in the Raman process \cite{Gros}. 
A second mechanism for a gap-induced mode may be 
a singlet bound state that splits off from the singlet-triplet 
excitation $\Delta_{SP}$ and is renormalized
due to dimer-dimer interaction  \cite{Uhrig}. Using RPA an estimate
of $\surd3\cdot\Delta_{SP}\approx$41-52~K was given as the energy 
for this
state \cite{Uhrig}. However, the spectral weight of this bound state in
Raman 
scattering is calculated to be negligible \cite{Gros}.

In Fig. 2 and 3 the Raman spectra of 
(Cu$_{1-x}$,Zn$_x$)GeO$_3$ and Cu(Ge$_{1-y}$,Si$_y$)O$_3$  
are presented for x=0, 0.018, 0.06 and y=0.022, 0.06, 0.1, 
respectively, at a) 5~K and b) 40~K. 
At first glance the spectra of (Cu$_{1-x}$,Zn$_x$)GeO$_3$
look very similar. Indeed, no big changes 
of the dominant phonons were observed. However, the gap-induced 
mode in (Cu$_{1-x}$,Zn$_x$)GeO$_3$ decreases strongly for 
x=0.018 and disappears for all x$>$0.018 (Fig.2a). For 
any Si-substitution investigated as shown in the upper panel of Fig. 3
the gap-induced mode is not visible. 
The Peierls-induced phonon modes 
at 105 and 370~cm$^{-1}$ and the peak of the two-magnon density 
of states for T$<$T$_{SP}$ are also reduced 
in intensity with increasing Zn- and Si-substitution. 
However, their appearence for higher x and y values
extends now to temperatures well above T$_{SP}$. 
This is most clearly seen in Fig. 3b for the highest Si-content y=0.1, 
but is also observed for smaller y or for Zn-substitution with x=0.06. 
This effect will be analyzed below in Fig. 6. The spinon continuum
observable 
above T$_{SP}$ and with a reduced intensity in the pure system 
below T$_{SP}$ is unchanged in magnitude in the case of Zn-
substitution 
up to x=0.06. With this value a maximum coherence length 
of the spinon excitations in the uniform state of 50~$\AA$
is calculated. 
Theoretical estimates \cite{Khomskii}
lead to coherence lenghts of the order
of $\xi_0\sim$(J/$\Delta_{SP}$)$\cdot$a=15-20~$\AA$, 
with the lattice parameter c=2.94$\AA$.
Additionally, the decrease of the continuum intensity 
is prohibited for x$\geq$0.035 due to a quenching of the spin-Peierls 
gap. 

The overall intensity of the continuum 
is modified for Si-substitution: We observe a gradual 
decrease of the continuum intensity with increasing y. 
The integrated intensity of the continuum for 
different x and y of both 
(Cu$_{1-x}$,Zn$_x$)GeO$_3$ and Cu(Ge$_{1-y}$,Si$_y$)O$_3$  
is presented in Fig. 4 as function of temperature. 
To facilitate a comparison of samples with different substitutions
we normalized the 
data to the phonon intensity at 593~cm$^{-1}$ for Zn-substitution or 
to 
the background in the frequency region 500-550~cm$^{-1}$ for the 
Si-substituted samples. The background as reference for 
Cu(Ge$_{1-y}$,Si$_y$)O$_3$  was chosen because of 
the non-negligible dependencies of the phonon intensity on 
Si-substitution. In Fig. 4a the continuum
intensity of (Cu$_{1-x}$,Zn$_x$)GeO$_3$ shows for x=0 and 0.018 
still a decrease below T$_{SP}$ of comparable size. For x$\geq$0.035 
this 
decrease is absent. In Cu(Ge$_{1-y}$,Si$_y$)O$_3$ the 
integrated intensity shows for y=0.022 and 0.06 a suppression 
of intensity below 40~K compared with y=0. This suppression 
results for  
y=0.1 in an overall smaller intensity of the 
continuum over the whole temperature region investigated
(5-300~K).
This effect on the frustration-induced scattering intensity 
is naturally attributed to a change of 
the relative frustration ($\beta-\alpha$) in $H_R$ and $H$ 
for the case of Si-substitution. 
As $\beta$ involves mainly excited orbital superexchange
which are probably not be changed by substitution we may
attribute the reduced scattering intensity of the 
continuum to an enlarged $\alpha$ (assuming $\alpha<\beta$).
This is supported by the strong sensitivity of $\alpha$
on pressure and the large variation of the lattice parameters
observed with Si-substitution \cite{Weiden3}.
Raman experiments with applied pressure were used to 
estimate an increase of $\alpha$ of 30$\%$ for a pressure
of 5~GPa \cite{Loos-press}. The lattice pressure 
observed in Si-substitution should lead to a similar 
effect and increase $\alpha$. In Zn-substituted samples
no large change of the lattice parameters exists 
\cite{Weiden1} and we do not observe a reduction of
intensity of the continuum for this substitution. 
It should be noted that the local character
of the Raman exchange process may be especially sensitive to 
local changes of the exchange path that are not visible in 
thermodynamic types of experiments. 
Indeed no shift of the broad maximum in the 
magnetic susceptibility was found in Si-substituted samples 
\cite{Weiden3}.
Therefore we propose neutron scattering  
as a key experiment to verify this result.

The gap-induced mode is replaced at low temperatures (5~K) 
by a gradual
increase of intensity for $\omega\rightarrow$0, both visible
for Si- and Zn-substitution (see Fig. 5). The quasi-elastic scattering 
of diffusive origin shows a continuous overall decrease of intensity 
with decreasing temperature 60 to 30~K.
Below 15~K, however, the intensity of the quasi-elastic scattering
increases again. 
This peculiar temperature dependence points to a nontrivial origin 
of the low temperature signal \cite{Martins,Hirsch,Schuettler},
i.e. not to scattering on static lattice defects
but to very low energy excitations of the spin system as they 
are predicted for doped spinons in substituted spin chains 
\cite{Martins}. 
It is remarkable that this intensity decreases again in the N\'eel state
of (Cu,Zn)GeO$_3$ at, e.g., T=1.8~K in Fig. 5a. 

In- and off-chain substitutions lead to anomalies in 
the Raman intensity that will be discussed here in the framework
of a dimensional crossover of CuGeO$_3$.
Both, the two-magnon cutoff 
peaking at 226~cm$^{-1}$ and the phonon mode at 370~cm$^{-1}$ 
are observed in substituted samples above the transition 
temperature of the pure compound. 
In Fig. 6 the normalized, integrated intensity of the 
370~cm$^{-1}$-phonon 
is shown for different Si- and Zn-substitutions. 
For the pure compound the 
onset of this mode is visible as a shoulder starting at T$_{SP}$ and 
increasing sharply below 12~K (see for comparison Fig. 1b on an 
enlarged scale). It reaches half its maximum intensity at
T$_{cross}$=11~K.
For Si-substituted samples T$_{cross}$ 
is determined to be 25~K(y=0.02), 35~K(y=0.06) and 
50~K(y=0.1). 
For Zn-substitution this mode is visible 
for x=0.06 at T$_{cross}$=23~K with 10\% intensity of the 
pure sample.  
With lower Zn-substitution levels the intensity is less
reduced. However, the shift of the crossover temperature 
above T$_{cross}$ of the pure compound is not as obvious
as in the case of Si-substitution. 
In Table I. a comparison of the characteristic temperatures of all
substituted
samples is given. The shift of T$_{cross}$ above T$_{SP}$ takes place 
for in-chain substitution with x$\ge$0.035 while for off-chain
substitution only y$\ge$0.005 is needed. 
This later value is determined by a linear
extrapolation of the characteristic temperatures in dependence on y.
The crossover is independent of the N\'eel temperature.

\section{DISCUSSION}
To understand the properties of CuGeO$_{3}$, the coupling of spin 
degrees of freedom to the lattice and the inter-chain interaction 
should be considered. This interplay is demonstrated in the 
behavior and dimensionality of the low-energy excitations 
observed in Raman scattering. Indeed, the two-magnon density of
states 
below 226~cm$^{-1}$ resembles 3D-excitations as it is fairly well 
described in a spin-wave approximation of a 3D Heisenberg System
\cite{Thomsen,Lemmens}.  The additional
phonon modes at 105, 370 and 820~cm$^{-1}$ undoubtedly have 3D 
character. 
The gap-induced mode at 30~cm$^{-1}$ and the spinon continuum, 
however, are attributed to 1D spinon excitations \cite{Muthu}. 
Hints towards excitations of mixed character exist in terms of a
Fano-lineshape
of the spin-Peierls-induced phonon at 105~cm$^{-1}$ and a high
energy
shoulder 
of the continuum above 400~cm$^{-1}$ that is not 
suppressed below T$_{SP}$ (see Fig. 1a). 
The low-energy maximum of this tail at 414~cm$^{-1}$ 
corresponds to the sum of the phonon at 370~cm$^{-1}$ and the 
spin-Peierls gap at the zone center, 
$\Delta_{\Gamma}$=44~cm$^{-1}$ \cite{Loosdrecht}.

The intensity of the Peierls-active phonon
at 370~cm$^{-1}$ and the two-magnon peak scale with 
each other in their dependence on temperature 
and, on the other hand, compete with the spinon continuum intensity.
As the key observation in our study, this leads to the following
conclusions:
The inter-chain interaction itself is important, but not sufficient to
explain
this rapid crossover in intensities. It should only lead to gradual
changes
of the spin dynamics below a 3D-coherence temperature 
T$_{3D}$=k$_B$J$_b\approx$15~K. However, the rotational
invariant
order of spin dimers formed at the spin-Peierls transition 
triggers this crossover
due to a rapid change of the s=1/2 spinons to s=1 excitations
\cite{Uhrig}.

Both inter-chain and long range
magnetoelastic interactions increase the k-phase space of 
the magnons from 1D to 3D and fix a long range ordered
arrangement
of dimers on the chains. This may result in a "confinement" 
of the spin-degrees of freedom and a decreasing spin-spin correlation
length
with increasing temperature.
A correct model should therefore include a coherence length
of 3D correlated spin dimers that is obviously temperature dependent.
Until now no such theory exists. Therefore we use arguments derived
on substituted spin ladders.

Recently, the theorectical study of non-magnetic impurities in spin
chains
and 
ladders gained interest as a first step towards
understanding correlation effects in doped quantum spin systems
\cite{Martins,Fukuyama2,Motome,Nagaosa,Iino}. 
In earlier Monte Carlo studies of the spin susceptibility in random
or disordered, non-dimerized spin chains 
a suppression of the long range AF correlations
and an enhancement of the spin-Peierls instability was found
\cite{Hirsch,Schuettler}. Recent theoretical investigations of strongly 
dimerized spin chains 
show that non-magnetic impurities create loose s=1/2 spins which
randomly 
introduce states within the magnetic excitation gap. 
These states should be observable as low energy excitations 
(doped spinons) that form a weakly 
dispersing impurity branch inside the gap. 
For realistic, smaller values of dimerization an 
effective spin s=1/2 is spread over several lattice spacings leading to
many-body s=1/2 states. These states enhance the spin-spin
correlation at short distances \cite{Martins,Motome}. In close relation
to these effects, studies of spin ladders propose a quantum phase
transition
from a gapped state to a quantum critical state
even at low impurity concentrations ($\approx$1\%) \cite{Motome}. 
Above this concentration a bulk spin state with no excitation gap and 
a long-range, algebraic decay of the spin-spin 
correlations are predicted for Zn-doped spin ladders \cite{Motome}.
The above mentioned effects were discussed in the substituted spin
ladder 
Sr(Cu,Zn)$_2$O$_3$ \cite{Martins,Azuma}.

The conclusion drawn from Raman experiments critically depends on
the 
interpretation of the two-magnon signal and the quasi-magnon
dispersion
seen in neutron scattering. It was demonstrated using neutron
scattering 
on CuGeO$_3$ \cite{Ain} 
that excitations with small k, close to k$_{AF}$=(0,1,0.5), are
dominantly 
influenced by the dimerization while the character of the magnons in
the 
middle of the Brillouin zone still resembles those of the spinons in
the uniform chain. 
Therefore excitations at low energy close to $\Delta_{SP}$ should
respond most strongly to an increase of temperature or impurity
content in the compound. 
Actually we see a strong broadening and a shift of the 
gap-induced mode close to T$_{SP}$ and its vanishing in
substituted samples while the two-magnon signal is only influenced
in intensity. 

In substituted samples we still see a similar behavior  
of the phonon and two-magnon intensities as shown in Figs. 2 and 3.
However the temperature range is shifted upwards.
The general existence of this scattering intensity 
and the shift of T$_{cross}$ 
above T$_{SP}$ of the pure compound may only be understood if an
increased
short range tendency towards dimerization exists at higher
temperature. 
This is seen in Monte Carlo studies on disordered spin chains as an 
enhanced four-spin susceptibility \cite{Hirsch}.
Similar results on substituted spin ladders 
point to an increased spin-spin correlation near an impurity 
\cite{Martins,Motome}.
In our case this enhanced correlation leads to a localization 
of a spin dimer at the impurity site. The crossover that was triggered
by the opening of the gap in the pure compound is therefore now
observed
at higher temperatures with T$_{cross}$ proportional to the impurity
content.
Si-substitution introduces more disorder compared with 
the local disruption of the spin chains using Zn-substitution.
Therefore the increase of T$_{cross}$ is more pronounced 
in this case. Our observations of a preferred spin-Peierls instability
against 
competing AF correlations in disordered spin chains 
are in good agreement with theory \cite{Hirsch}.

As shown in the previous chapters the spin-Peierls active 
modes react in various ways on substitution. This difference is in part
originated in the different coherence length of the excitations
involved.
The 1D spinon continuum is observed both in pure and substituted
samples.
By the opening of the spin-Peierls gap this intensity is reduced
due to the localizing, pair-binding effect of the excitation gap.  
No drop in the continuum intensity is observed for sufficiently 
large Zn- (x$>$0.03) or Si-substitution (y$>$0.02). 
The survival of the continuum, independent 
of substitution, clearly stresses the short-range character of 
these excitations in the dimerized phase. 

In general it is quite striking that only a small amount of substitution
is required to completely suppress the spin-Peierls transition, 
independent of the substitution site. The important point seems to be
that 
the system is disturbed by substitution. 
The appearence of N\'eel ordering in a limited concentration range
starting
above $\approx$1\% for both substituted CuGeO$_3$ and the spin ladder
system
SrCu$_2$O$_3$ points to a similar instability. A proposed quantum 
phase
transition \cite{Martins,Motome} in weakly substituted 
spin ladders is a promising concept
to describe this behavoir as it also gives hints towards the
disappearence
of T$_N$ at higher concentrations. Here the long ranged resonating
valence bond state gets 
energetically unfavorable again.
To interpret the appearence and consequent disappearence of T$_N$ 
simply due to inter-chain 
coupling and a following frustration of the 3D ordered spin system
would
not be appropriate as the pure compound is already strongly
frustrated. 
The absence of any additional magnetic scattering for T$\leq$T$_N$
in CuGeO$_3$ is quite striking. This means
that on the local scale of exchange scattering 
no additional scattering channel 
opens up going from the regime of short range AF fluctuations with
coexisting
local singlet formation to the truly ordered 3D AF state. However,
the quasi-elastic scattering observed at low temperatures in substituted
samples attributed to low lying defect-induced spin excitations
disappears
in the N\'eel state.


\section{CONCLUSIONS}

Using Raman scattering the effect of in-chain and off-chain
substitutions
on spin fluctuations and spin-Peierls active modes was investigated.
The observed scattering modes respond differently on 
changes of the temperature and on substitutions. 
The gap-induced mode showing up 
below T$_{SP}$ strongly decreases in intensity upon Zn- and
Si-substitution.
Additionally, the drastic decrease of the integrated spinon continuum 
below T$_{SP}$ is suppressed in substituted samples due to a quench
of the spin-Peierls gap. This effect does not depend on 
the kind of defect and is in agreement with magnetic 
susceptibility measurements on 
substituted CuGeO$_{3}$ \cite{Weiden2}.
However, the decrease of the spinon continuum that was observed 
above T$_{SP}$ only for 
Si-substitution is tentatively attributed to lattice pressure-induced 
changes of the frustration. Si-substitution leads to the largest changes
of the lattice parameters of all investigated substitutions
\cite{Weiden1}.

The modes of 3D origin, i.e. the spin-Peierls 
induced phonon at 370~cm$^{-1}$ 
and the two-magnon cutoff at 226~cm$^{-1}$ compete in the pure
samples with the 1D spinon continuum. 
A crossover temperature is proposed that marks a transition from 
1D to 3D behavior of the coupled spin chains. 
This crossover is triggered by a 3D coherent singlet state
arising due to the inter-chain interaction.
In the substitution studies presented here the crossover temperature is
shifted to higher temperatures due to enhanced spin-spin correlation at
the impurity site. Additional antiferromagnetic 
low lying spin excitations are induced by the 
substitution. These observations are in good agreement with recent 
theoretical calculations of the spin response and behavior of 
the spin-spin correlation length in substituted spin ladders. 
This supports our idea of a general
similarity between defect states in weakly coupled
spin chains close to a spin-Peierls transition and 
spin ladders where the dimer formation is favored by topology.

Acknowledgment: This work was supported by DFG, through SFB
341, 
SFB 252 and by BMBF 13N6586/8. We thank 
V.N. Muthukumar, W. Brenig and P.H.M van Loosdrecht
for extensive and helpfull discussion. 
%
%

\newpage
\begin{table}
\caption{Investigated single crystals
(Cu$_{1-x}$,Zn$_x$)(Ge$_{1-y}$,Si$_y$)O$_3$
with their T$_{SP}$ and T$_N$ determined by SQUID magnetometry. 
T$_{cross}$ is derived by Raman scattering and marks a crossover 
from 3D to 1D behavior in the spin excitations.
A dash (-) denotes no observation of a phase transition.}
\begin{center}
\begin{tabular}{|c|c|c|c|c|}
x (Zn)&y (Si)&T$_{SP}$/K&T$_N$/K &T$_{cross}$/K\\
\hline \hline
0&0&14.3&-&11\\
\hline
0.018&&12.8&-&11\\
0.035&&11&2.5&11\\
0.06&&-&4&23\\
\hline
&0.022 & -&5&25\\
&0.04&-&4.5&30\\
&0.06&-&2.5&35\\
&0.1&-&-&50\\
\end{tabular}
\end{center}
\end{table}
{\bf Figure Captions}\\

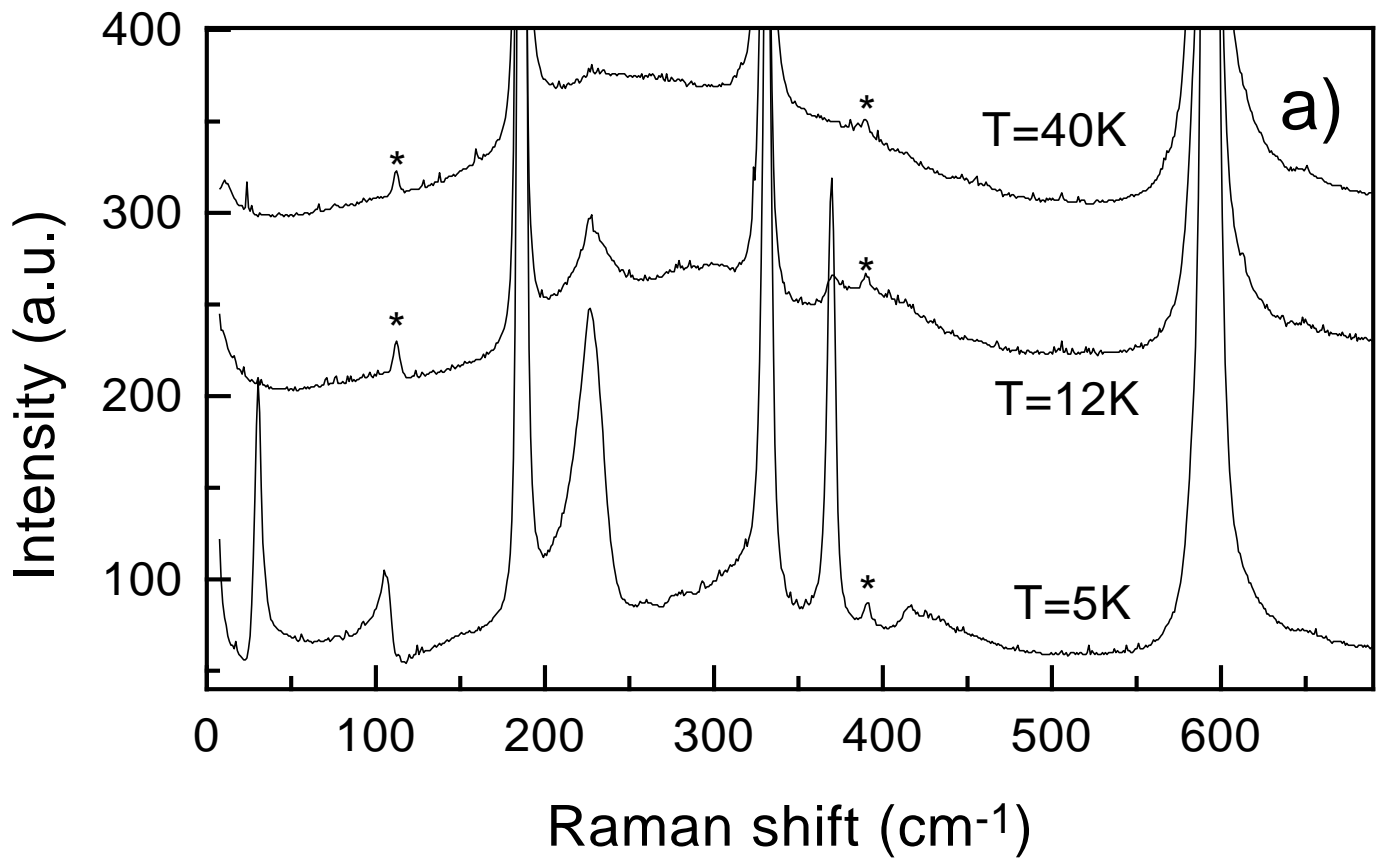
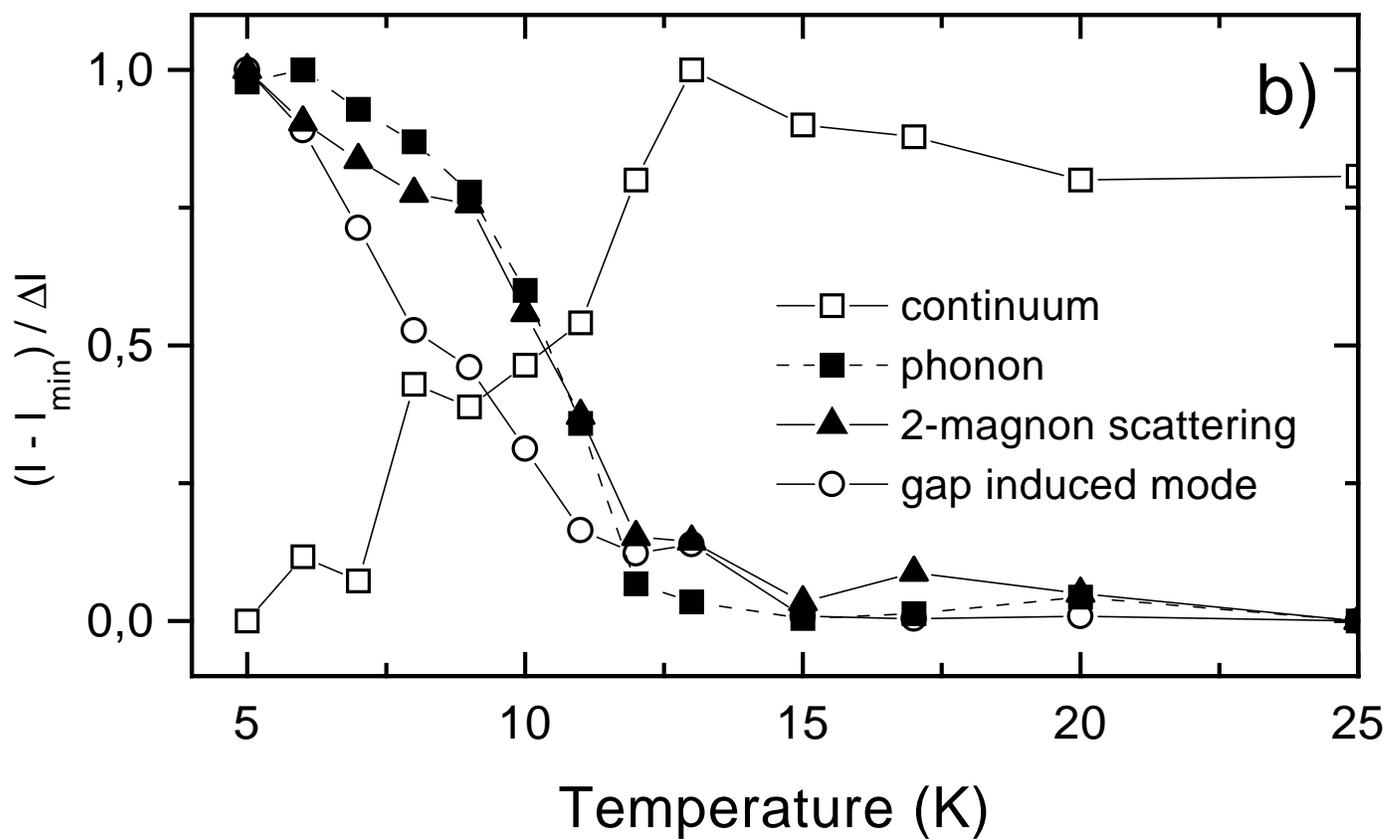
Fig. 1. Raman scattering intensity of CuGeO$_3$, a) 
for temperatures between 5 and 40~K.
Additional symmetry forbidden modes due to a slight misalignment
of the sample are marked by an asterisk.
b) Temperature dependence of the 
renormalized intensity of the spinon continuum, 
the Peierls-transition-induced phonon at 370~cm$^{-1}$, the  
two-magnon density of states cutoff at 226~cm$^{-1}$ 
and the gap-induced mode at 30~cm$^{-1}$. \\

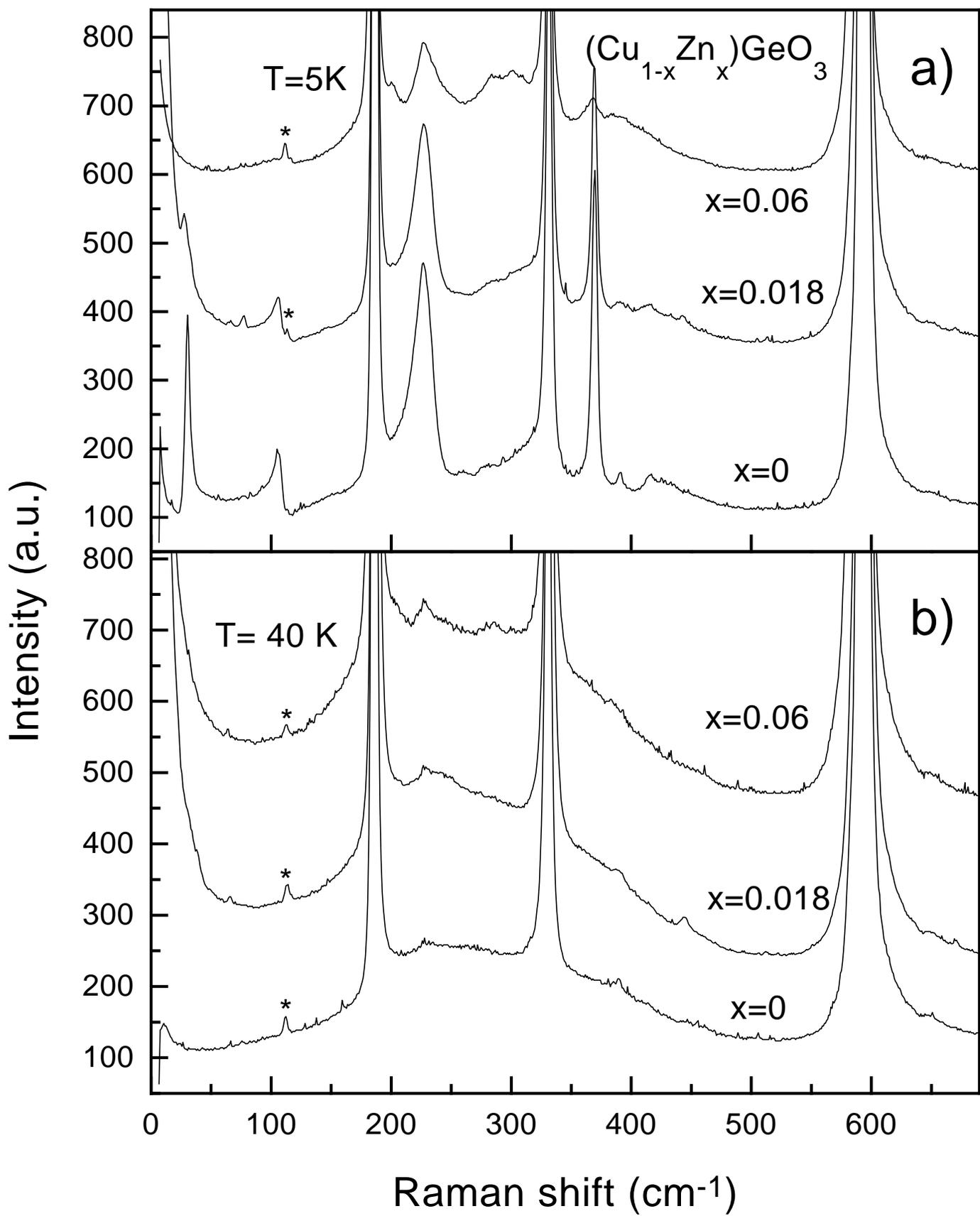
Fig. 2. Scattering intensity of Cu$_{1-x}$Zn$_x$GeO$_3$ 
with x=0, 0.018, 0.035 and 0.06 at a) 5 and b) 40~K. 
The curves are shifted by a subsequent addition of 200 a.u. 
Additional symmetry forbidden modes due to a slight misalignment
of the sample are marked by an asterisk.\\

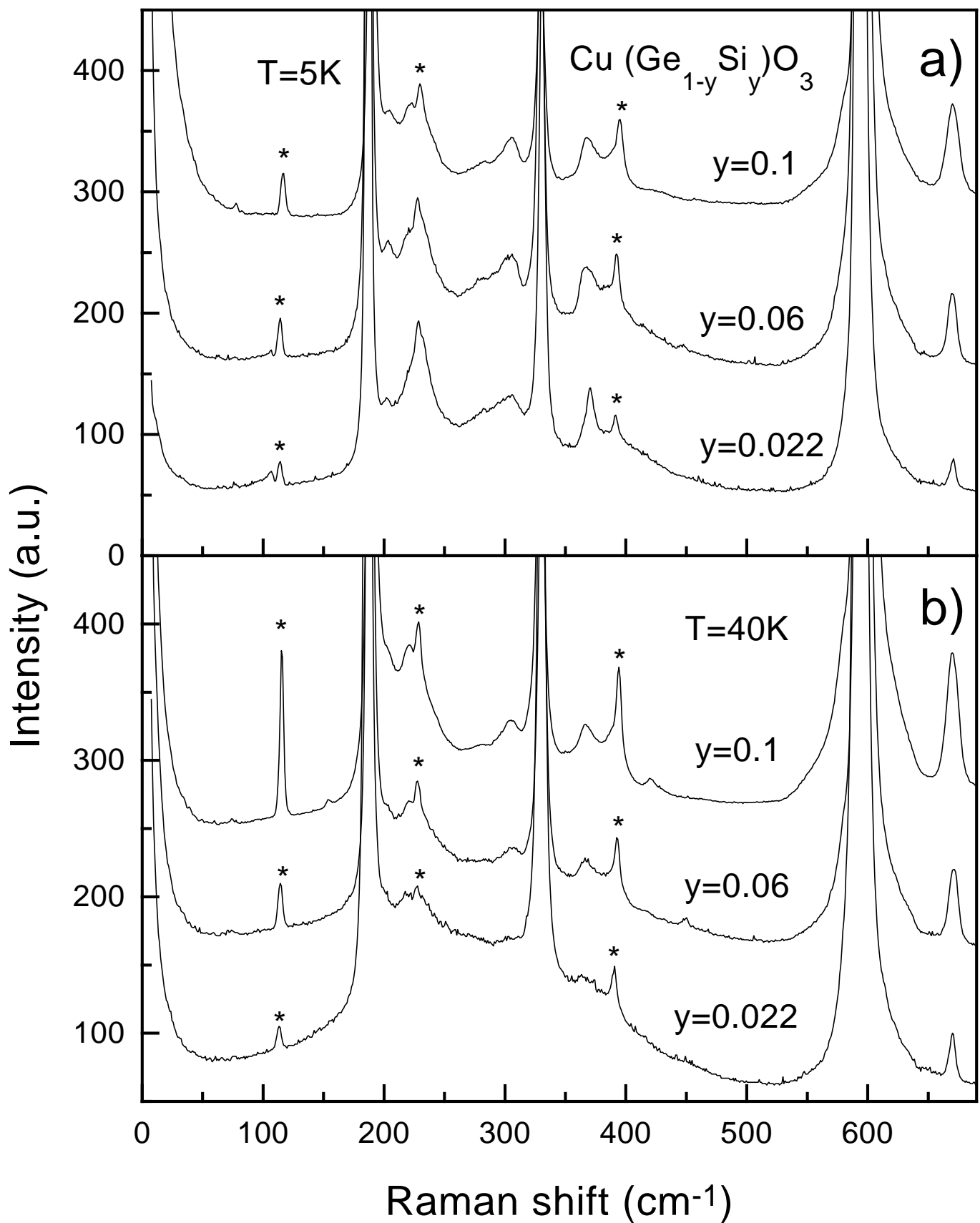
Fig. 3. Scattering intensity of Cu(Ge$_{1-y}$,Si$_y$)O$_3$ 
with y=0, 0.022, 0.06, 0.1 at a) 5 and b) 40~K.
The curves are shifted by a subsequent addition of 100 a.u.
Symmetry forbidden modes of the pure compound that are 
induced by Si-substitution are marked by an asterisk. The phonon at
690~cm$^{-1}$ is a defect mode.\\

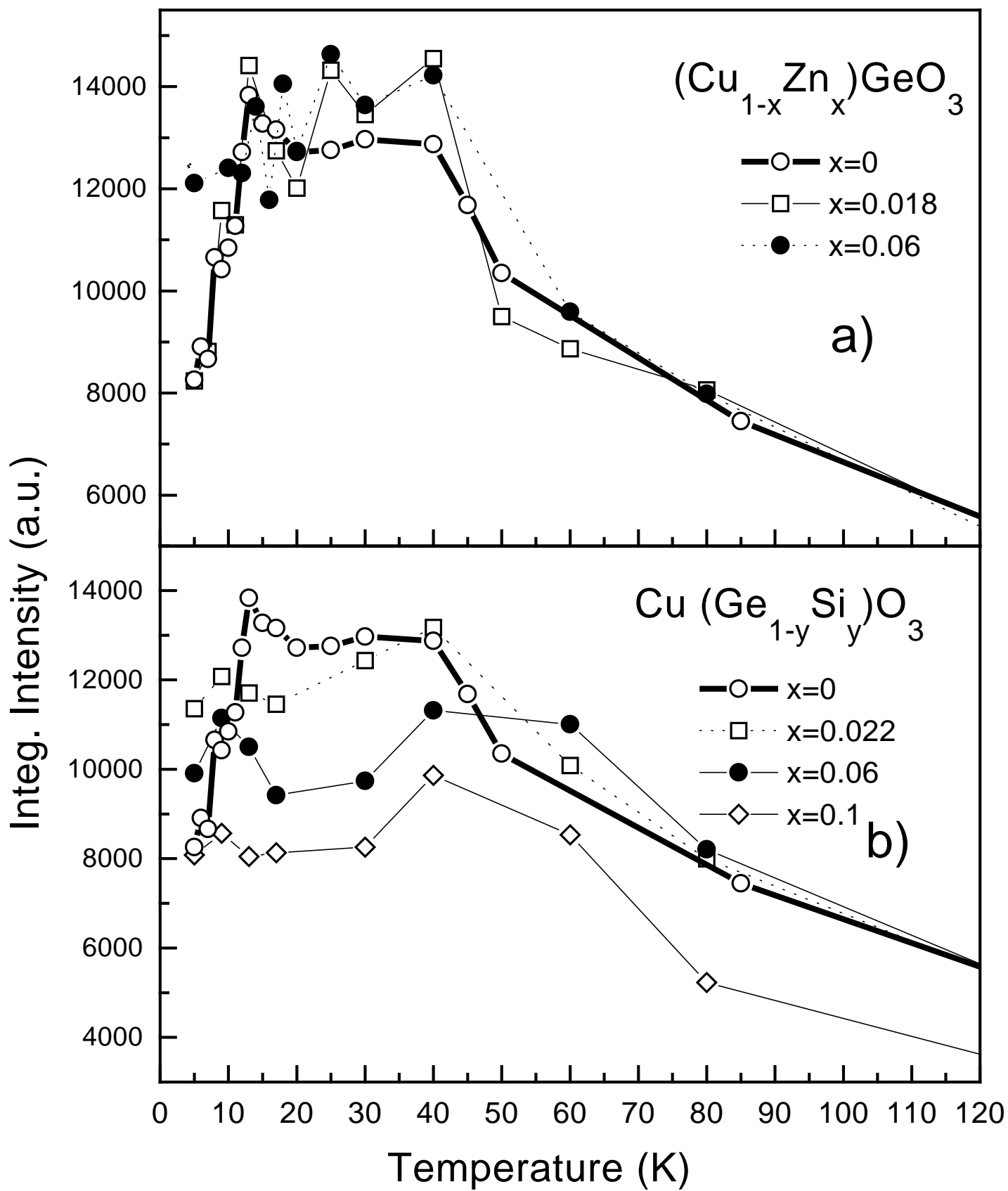
Fig. 4. Integrated intensity of the spinon continuum between 
100-500~cm$^{-1}$ for a) Cu$_{1-x}$Zn$_x$GeO$_3$ with 
x=0, 0.018, 0.06 and b) Cu(Ge$_{1-y}$,Si$_y$)O$_3$ with 
y=0, 0.022, 0.006, 0.1 in dependence of temperature. 
A linear background, phonons and additional modes were subtracted 
from the data prior to integration. The data on
Cu$_{1-x}$Zn$_x$GeO$_3$ 
was normalized to the phonon at 593~cm$^{-1}$ while the data 
on Cu(Ge$_{1-y}$,Si$_y$)O$_3$ was normalized on the 
background between 500-550~cm$^{-1}$ to facilitate the comparison 
between different x, y, respectively. \\

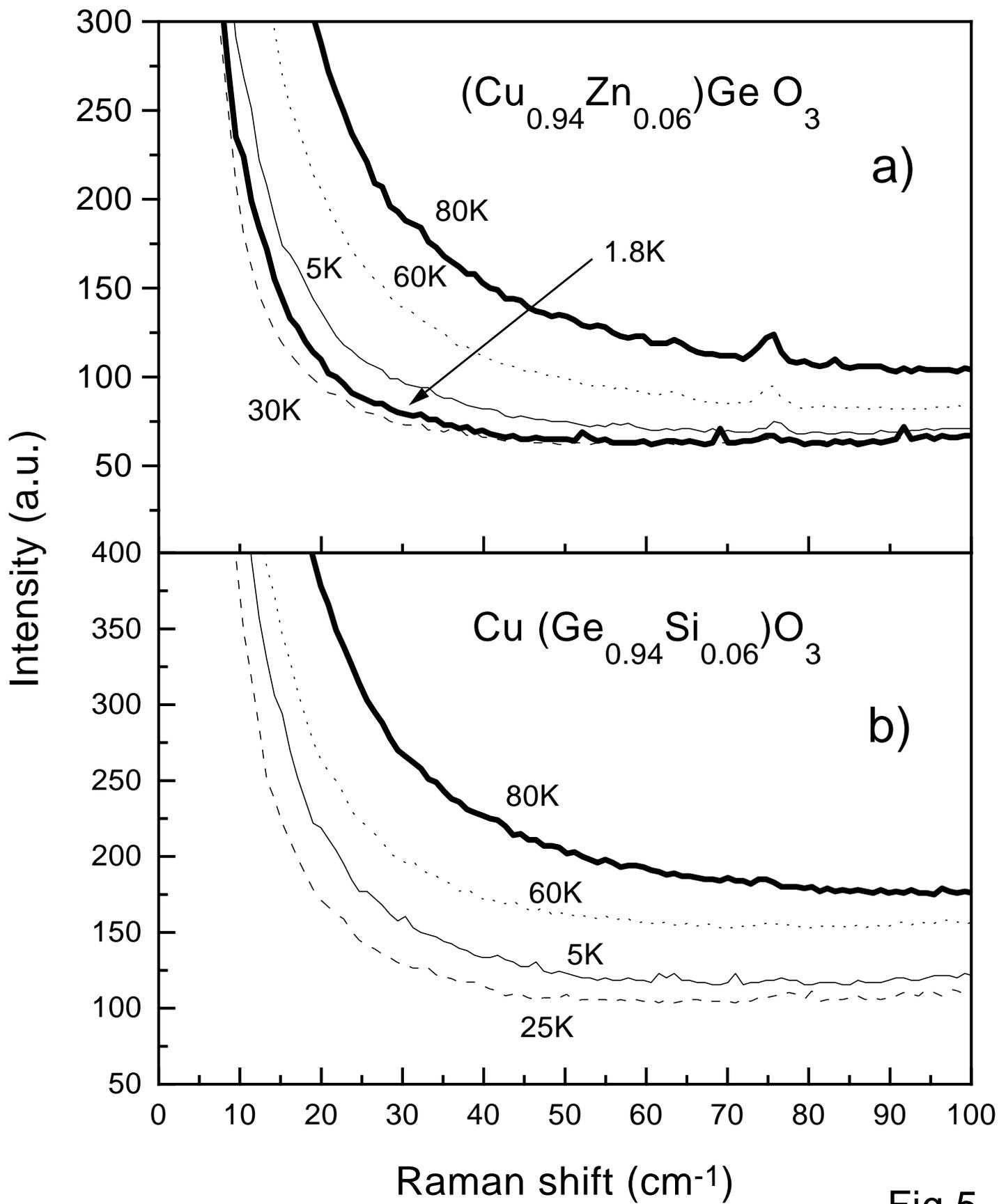
Fig. 5. Quasielastic scattering in a) Cu$_{0.94}$Zn$_{0.06}$GeO$_3$
and b) Cu(Ge$_{0.94}$,Si$_{0.06}$)O$_3$ for different temperatures.
The signal at T=30 and 25~K remains from straylight. The intensity
at T=5~K is above this level.\\ 

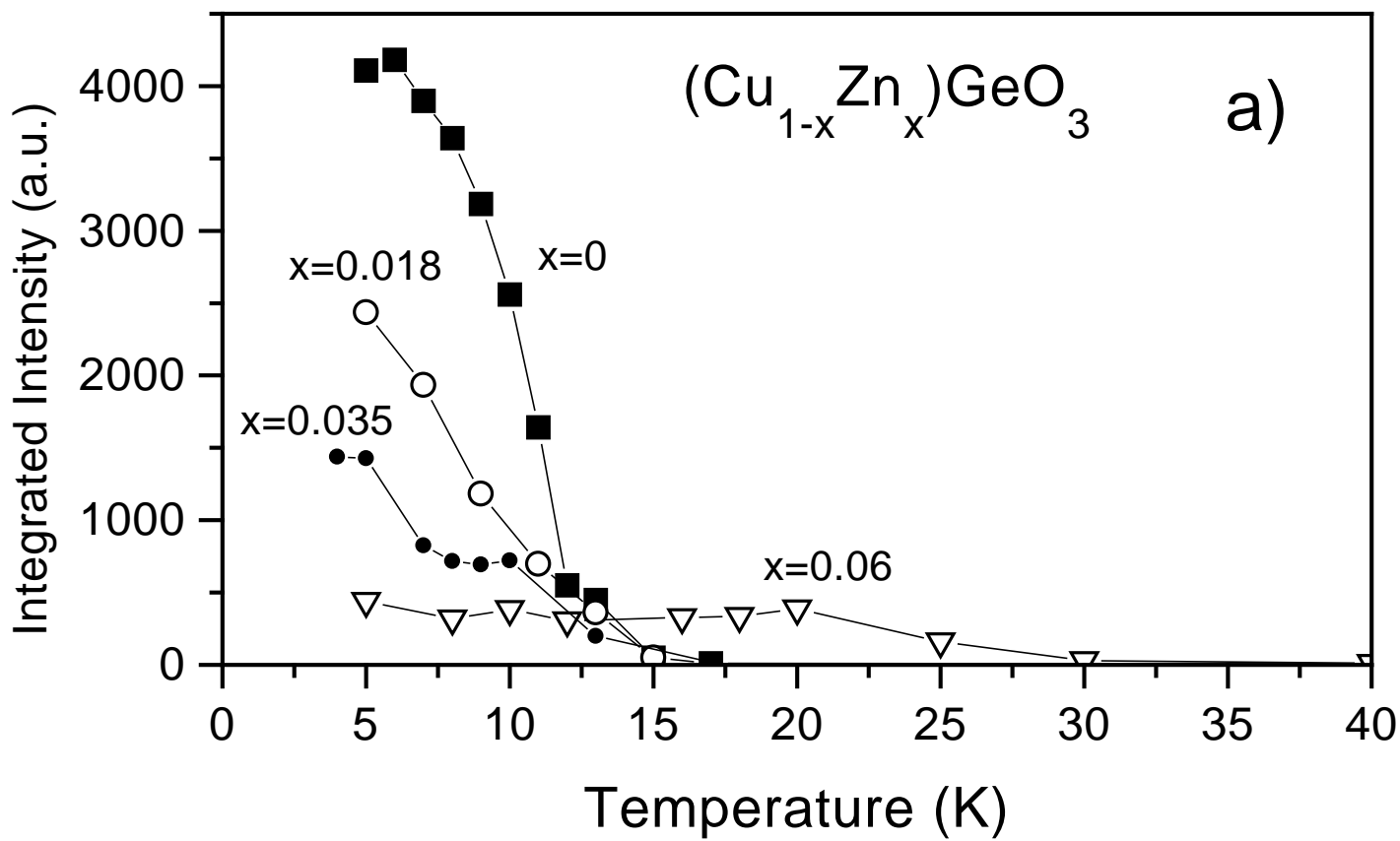
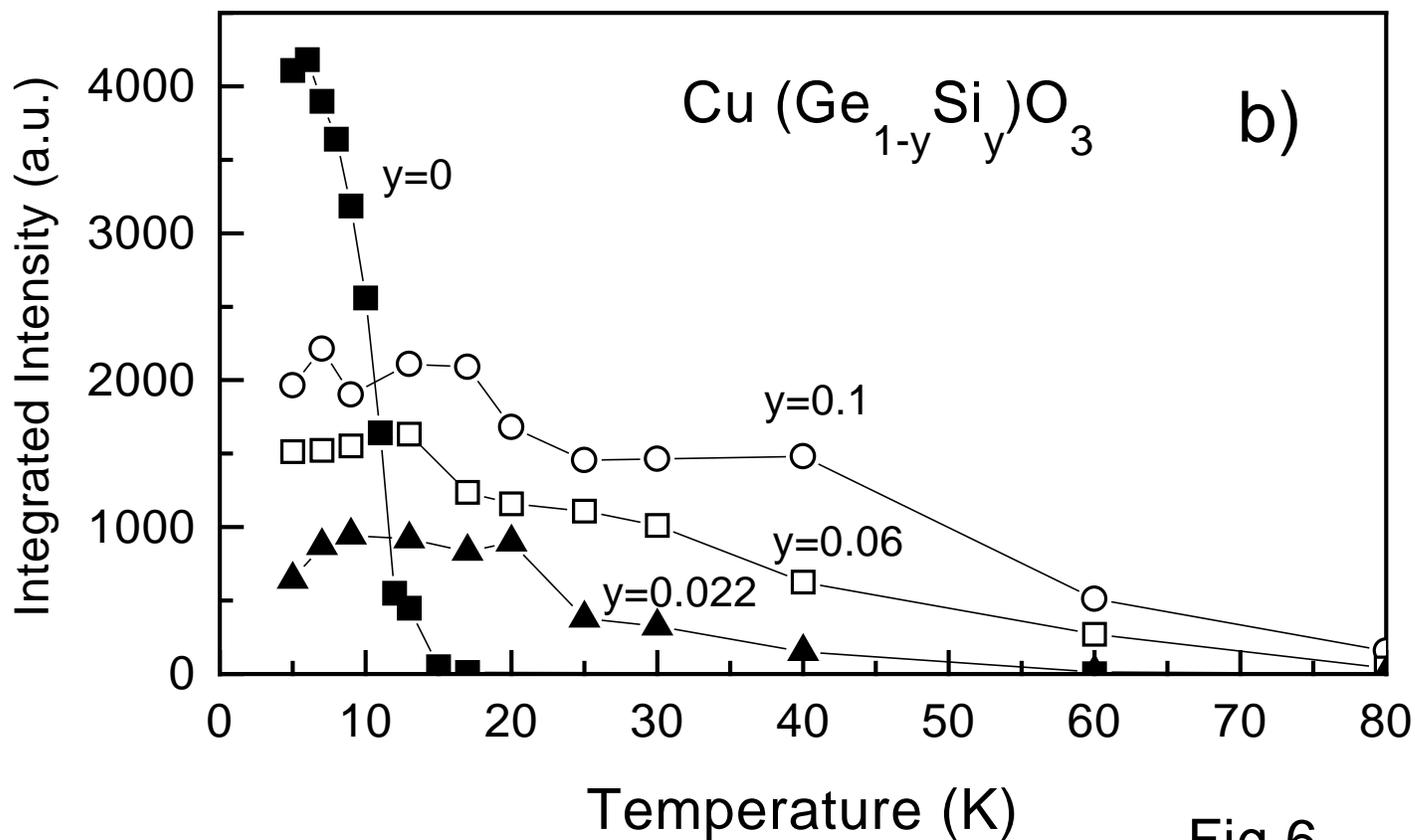
Fig. 6. Normalized integrated intensity of the 
spin-Peierls-induced phonon mode 
at 370~cm$^{-1}$ for Cu$_{1-x}$Zn$_x$GeO$_3$ and
Cu(Ge$_{1-y}$,Si$_y$)O$_3$. 
T$_{cross}$ is defined from this data 
as the temperature of half maximum intensity of the phonon.\\


\begin{thebibliography}{99}

\bibitem{Mueller} G. M\"uller, 
	H. Thomas, H. Beck and J.C. Bonner, 
	Phys. Rev. B{\bf 24}, 1429 (1981).

\bibitem{Cloizeau} J. des Cloizeaux and J.J. Pearson, 
	Phys. Rev. {\bf 128}, 2131 (1962).

\bibitem{KCuF3} D.A. Tennant, 
	R.A. Cowley, S.E. Nagler and A.M. Tsvelik,
	Phys. Rev. B{\bf 52}, 13 368 (1995).

\bibitem{Arai} M. Arai, M. Fujita, M. Motokawa, 
	J. Akimitsu, S.M. Bennington,
	Phys. Rev. Lett {\bf 77}, 3649 (1996). 

\bibitem{Uhrig} G.S. Uhrig, H.J. Schulz, 
	Phys. Rev. B{\bf 54}, R9624 (1996);
        A. Fledderjohann, C. Gros,
	to be published in Europhys. Lett.
        (Sissa preprint cond-mat/9612013).

\bibitem{Hase} M. Hase, I. Terasaki, K. Uchinokura,
	Phys. Rev. Lett. {\bf 70}, 3651 (1993).

\bibitem{Nishi} M. Nishi, 
	O. Fujita and J. Akimitsu,
	Phys. Rev. B{\bf 50}, 6508 (1994).

\bibitem{Raman} H. Kuroe {\it et al.},
	Phys. Rev. B{\bf 50}, 16 468 (1994).

\bibitem{Boucher} J.P. Boucher, L.P. Regnault, 
	to be published in Journal de Physique.

\bibitem{Castilla} G. Castilla,
	S. Chakravarty and V.J. Emery, 
	Phys. Rev. Lett. {\bf 75}, 1823 (1995).

\bibitem{Buech} B. B\"uchner, U. Ammerahl, T. Lorenz, 
	W. Brenig, G. Dhalenne, A. Revcolevschi,
	Phys. Rev. Lett. {\bf 77}, 1624 (1996).

\bibitem{Braden} M. Braden, G. Wilkenorf, 
	J. Lorenzana, M. Ain, G.J. McIntyre, 
	M. Behruzi, G. Heeger, G. Dhalenn, A. Revcolevschi, 
	Phys. Rev. B{\bf 54}, 1105 (1996). 

\bibitem{Khomskii} D. Khomskii, W. Geertsma, M. Mostovoy, 
	Czech. Journal of Physics {\bf 46}, 32-39 (1996).

\bibitem{Chitra} R. Chitra, et al.,
	Phys. Rev. B{\bf 52}, 6581 (1995).

\bibitem{Martin} M. C. Martin, 
	G. Shirane, Y. Fujii, M. Nishi, O. Fujita, 
	J. Akimitsu, M. Hase, K. Uchinokura, 
	Phys. Rev. B{\bf 53}, R14713 (1996).
	
\bibitem{Ain} M. Ain,
	J.E. Lorenzo, L.P. Regnault, G. Dhalenne, 
	A. Revcolevschi,  Th. Jolicoeur, 
	{\it ``Double gap and solitonic excitations in the 
	spin-Peierls chain CuGeO$_3$创}, unpublished.

\bibitem{Regn2} L.P. Regnault, J.P. Renard, 
	G. Dhalenne, A. Revcolevschi,
	Europhys. Lett. {\bf 32}, 579 (1995).

\bibitem{Hase2} M. Hase, K. Uchinokura, R.J. Birgeneau
	K. Hirota, G. Shirane,
	J. Phys. Soc. Japn. {\bf 65}, 1392 (1996). 

\bibitem{Fukuyama} H. Fukuyama, T. Hamitomoto, M. Saito,
	J. Phys. Soc. Japn. {\bf 65}, 1182 (1996) 
	
\bibitem{Loosdrecht} P.H.M. van Loosdrecht {\it et al.},
	Phys. Rev. Lett. {\bf 76}, 311 (1996) and\\
	P.H.M. van Loosdrecht, J.P. Boucher, S. Huant, 
	G. Martinez, G. Dhalenne, A. Revcolevschi,  
	Proceedings of SCES 1996, Z\"urich, 
	to be published in 
	Physica {\bf B} (1997).  

\bibitem{Thomsen} I. Loa, S. Gronemeyer, C. Thomsen, 
	Solid State Commun., {\bf 99}, 4, 231-235 (1996).

\bibitem{Lemmens} P. Lemmens, 
	B. Eisener, M. Brinkmann, L.V. Gasparov, 
	G. G\"untherodt, P.v. Dongen, M. Weiden, 
	W. Richter, C. Geibel, F. Steglich, 
	Physica {\bf B223\&224}, 535-537 (1996) and \\
	P. Lemmens, M. Udagawa, M. Fischer, G. G\"untheroth, 
	M. Weiden, W. Richter, C. Geibel, F. Steglich, 
	Czechoslovak J. Phys. {\bf 46}, 1979-1980 (1996) and \\
	M. Fischer, P. Lemmens, G. G\"untherodt, M. Weiden, 
	W. Richter, C. Geibel, F. Steglich,
	{\it ``Substitution effects on spin fluctuations 
	in the spin-Peierls compound CuGeO$_3$创}, 
	Proceedings of SCES 1996, Z\"urich, 
	to be published in Physica {\bf B}, (1997).

\bibitem{Muthu} V.N. Muthukumar, 
	C. Gros, W. Wenzel, R. Valent\'i, P. Lemmens, B. Eisener, 
	G. G\"untherodt, M. Weiden, C. Geibel, F. Steglich,
	Phys. Rev. B{\bf 54}, R9635 (1996) and \\
	V.N. Muthukumar, C. Gros, W. Wenzel, 
	R. Valent\'i, M. Weiden, C. Geibel, F. Steglich, 
	P. Lemmens, M. Fischer, G. G\"untherodt, 
	{\it ``The J$_1$ - J$_2$ model revisited: 
	Phenomenology of CuGeO$_3$创},
	to be published in Phys. Rev. B. 

\bibitem{Fleury} P.A. Fleury and R. Loudon,
	Phys. Rev. {\bf 166}, 514 (1967).

\bibitem{Weiden1} M. Weiden, 
	W. Richter, C. Geibel, F. Steglich, 
	P. Lemmens, B. Eisener, M. Brinkmann, 
	G. G\"untherodt,  
	{\it ``Doping Effects in CuGeO$_3$创}, 
	accepted for publication in Physica {\bf B} (1997). 

\bibitem{Weiden2} M. Weiden, W. Richter, R. Hauptmann, 
	C. Geibel, F. Steglich, 
	{\it ``Universal Phase Diagram of Doped 
	Spin-Peierls-System CuGeO$_3$创}, 
	unpublished.

\bibitem{Weiden3} M. Weiden, W. Richter, R. Hauptmann, 
	C. Geibel, P. Hellmann, M. K\"oppen, F. Steglich
	N. Weiden, M. Fischer, P. Lemmens, G. G\"untherodt,
	A. Krimmel, G. Nieva, 
	{\it ``Phase Diagram of CuGe$_{1-x}$Si$_x$O$_3$}, 
	to be published in PRB. 

\bibitem{Renard} J.P. Renard, K. deDang, P. Veillet, 
	G. Dhalenne, A. Revcolevschi, L.P. Regnault, 
	Europhys. Lett. {\bf 30} (8), 475 (1995).

\bibitem{Fischer} M. Fischer, P. Lemmens, G. G\"untherodt, 
	M. Weiden, C. Geibel, F. Steglich, 
	unpublished.

\bibitem{Cottam} M.G. Cottam, D.J. Lockwood, 
	{ \it ``Light scattering in magnetic solids创}, 
	(Wiley-Interscience Publ., 1986).

\bibitem{Gros} C. Gros, W. Wenzel,
	A. Fledderjohann, P. Lemmens,
	M. Fischer, G. G\"untherodt,
	M. Weiden, C. Geibel, F. Steglich, 
        submitted to Phys. Rev. B.

\bibitem{Loos-press}  P.H.M. van Loosdrecht,
	J. Zeman, G. Martinez, G. Dhalenne, A. Revcolevschi,  
	Phys. Rev. Lett., Jan. 1997.

\bibitem{Martins}G.B. Martins, E. Dagotto, 
	{\it Rapid suppression of the Spin Gap 
	in Zn-doped CuGeO$_3$ and SrCu$_2$O$_3$}, 
	Babbage preprint, cond-mat/9605069,
	subm. to Phys. Rev. B.

\bibitem{Hirsch} J.E. Hirsch, R. Kariotis,
	Phys. Rev. B{\bf 32}, 7320 (1985).

\bibitem{Schuettler} H.-B. Sch\"uttler, D.J. Scalapino, P.M. Grant,
	Phys. Rev. B{\bf 35}, 3461 (1987).

\bibitem{Fukuyama2} H. Fukuyama, N. Nagaosa, 
	M. Saito, T. Tanimoto,
	J. Phys. Soc. Japn. {\bf 65}, 2377 (1996). 

\bibitem{Motome} Y. Motome, N. Katoh, 
	N. Furukawa, M. Imada, 
	J. Phys. Soc. Japn. {\bf 65}, 1949 (1996). 

\bibitem{Nagaosa} N. Nagaosa, 
	A. Furusaki, M. Sigrist, H. Fukuyama, 
	{\it ``Non-magnetic Impurities in Spin-Gap Systems创}, 
	Babbage preprint, cond-mat/9609016, 
	to be published in J. Phys. Soc. Jap.

\bibitem{Iino} Y. Iino, M. Imada, 
	{\it ``Effects of Nonmagnetic Impurity 
	Doping on Spin Ladder Systems创}, 
	Babbage preprint, cond-mat/9609038,
	subm. to J. Phys. Soc. Jap.

\bibitem{Azuma} M. Azuma, Y. Fujishiro, M. Takano, 
	T. Ishida, K. Okuda, M. Nohara, T. Takagi, 
	unpublished.
  
\end{thebibliography}
\end{document}